\documentclass[%
 reprint,
superscriptaddress,
 amsmath,amssymb,
 aps,
prl,
]{revtex4-2}

\usepackage{graphicx}
\usepackage{dcolumn}
\usepackage{bm}
\usepackage{svg}
\usepackage[T1]{fontenc}
\usepackage{siunitx}

\begin{document}

\preprint{APS/123-QED}

\title{Enhancing exotic quantum fluctuations in a strongly entangled cavity BEC system}
\author{Leon Mixa}
\email{lmixa@physnet.uni-hamburg.de}
\affiliation{I. Institut für Theoretische Physik, Universität Hamburg, Notkestraße 9, 22607 Hamburg, Germany}
\author{Hans Ke{\ss}ler}
\affiliation{Zentrum für Optische Quantentechnologien and Institut für Laser-Physik, Universität Hamburg, 22761 Hamburg, Germany}
\author{Andreas Hemmerich}
\affiliation{Zentrum für Optische Quantentechnologien and Institut für Laser-Physik, Universität Hamburg, 22761 Hamburg, Germany}
\affiliation{The Hamburg Center for Ultrafast Imaging, Luruper Chaussee 149, 22761 Hamburg, Germany}
\author{Michael Thorwart}
\affiliation{I. Institut für Theoretische Physik, Universität Hamburg, Notkestraße 9, 22607 Hamburg, Germany}
\affiliation{The Hamburg Center for Ultrafast Imaging, Luruper Chaussee 149, 22761 Hamburg, Germany}

\begin{abstract}
We show that the strong coupling of a quantum light field and correlated quantum matter 
induces exotic quantum fluctuations in the matter sector. 
We determine their spectral characteristics and reveal the impact of the atomic s-wave scattering.  In particular, we derive the dissipative Landau and Beliaev processes from the microscopic Hamiltonian using imaginary time path integrals. By this, their strongly sub-Ohmic nature is revealed analytically. A competition between damping and antidamping channels is uncovered. 
Their intricate influence on physical observables is quantified analytically and the Stokes shift of the critical point is determined. 
This illustrates the tunability of the quantum matter fluctuations by exploiting strong light-matter coupling.
\end{abstract}

\maketitle

When an ultracold atomic gas is placed in an optical cavity, strong light-matter interaction can give rise to a rich landscape of quantum phases \cite{baumann2010,brennecke2013,klinder2015dynamical,klinder2015observation,zhang2021observation}.  A prominent example is the second-order non-equilibrium Dicke quantum phase transition \cite{emary2003quantum}.  It can be observed by strongly coupling an atomic Bose-Einstein condensate (BEC) to a cavity, pumped with a retro-reflected laser beam, which is adjusted perpendicularly to the cavity axis. Above a critical pump strength, the atomic density, in a spontaneous ${\displaystyle \mathbb {Z} _{2}}$ symmetry breaking process, switches from a homogeneous state to one of two possible density gratings, which resemble the black or white fields of a checkerboard lattice. Each density grating fulfills the Bragg condition to maximize coherent scattering  \cite{domokos2002collective} of photons from the pump field into the cavity and vice versa. The resulting interference between the pump and intra-cavity light fields leads to a two-dimensional light shift potential, which stabilizes the prevailing density grating.

A two-mode description of the atomic field operator can explain this phenomenon \cite{nagy2008self}. 
The photon scattering couples the zero-momentum state of the condensate to a symmetric superposition of four finite-momentum condensate states $|\pm k, \pm k \rangle$, which carry two quanta of photon momentum $\hbar k$, one along the cavity, and one along the pump axis \cite{baumann2010}.
Eventually, condensate fluctuations trigger the transition. At zero temperature, quantum fluctuations determine the fate of the condensate \cite{nagy2008self}. 
The scattering of the pump photons creates polaritons that involve both the collective oscillations of the condensate and the intracavity field \cite{nagy2008self}.
The excitation spectrum provides a soft mode whose frequency vanishes as the critical point is approached \cite{nagy2008self}.

Even if dissipation by cavity loss is included,  the criticality survives \cite{nagy2011critical}, yet with the critical exponent modified from $1/2$ to $1$.
Due to the quantum correlations between the cavity photons and the atoms, this dissipation channel enables non-destructive measurement of the atomic state \cite{mekhov2007cavity,mekhov2007light,mekhov2007probing} by detecting the out-going photons. Hence, fluctuations in the matter and light sectors can be studied in situ. 

Commonly, 
damping of the excited atomic state is introduced phenomenologically \cite{brennecke2013}. On the microscopic level,  damping of quasiparticles in a superfluid occurs as Landau \cite{guilleumas1999temperature,jackson2002accidental,guilleumas2003landau,jackson2003landau,tsuchiya2005landau} and Beliaev damping \cite{hodby2001experimental,kagan2001damping,katz2002beliaev}.
Within a Born-Markov approximation, the damping rate of the polariton soft mode is known \cite{konya2014damping}.
Landau damping vanishes at zero temperature and grows monotonously with a vanishing soft mode frequency.
Beliaev damping peaks when the polariton decays into two phonons close to the edge of the Brillouin zone \cite{konya2014damping} which induces a non-analytical point in the fluctuation spectrum of the condensate \cite{konya2018nonequilibrium}.  The Born-Markov approximation breaks down when the two significant poles of the polariton Green function form an avoided crossing \cite{konya2014damping}.  Indeed, phenomenological sub-Ohmic baths do modify the critical exponent \cite{nagy2015nonequilibrium}. Sub-Ohmic reservoirs induce strong non-Markovian dynamics and come along with exotic quantum-colored noise features \cite{Weiss}. They originate due to charge noise in superconducting qubits \cite{qubits}, quantum dots \cite{qudots}, quantum impurity systems \cite{quimp1,quimp1}, and nanomechanical oscillators \cite{nanomech} and rule the ultraslow glass dynamics \cite{glass}. Sub-Ohmic dissipative quantum dynamics is notoriously difficult to calculate \cite{subOhmic}.

In this work, we formulate a generalized  Bogoliubov theory to microscopically derive the colored spectra of the quantum statistical fluctuations. They show exotic, strongly sub-Ohmic features, yielding significant non-Markovian dynamics. An imaginary time path integral allows us to describe the formation of the dissipative two-mode polariton in an exact analytic fashion. Finally, we determine the influence of damping on the physical observables, illustrating thereby the possibility to control and enhance the quantum fluctuations in the system.

{\em Model. --} We consider the internal state transition $\omega_A$ of $N_C$ atoms which is transversely pumped by a laser with the frequency $\omega_P$  far detuned by $\Delta_A = \omega_P - \omega_A$.
The pump mode function is $h(\bm{r}) = h_P \cos{ky}$ with the Rabi frequency $h_P$. 
The single mode cavity with frequency $\omega_C$ hosting the atoms is described by the cavity photon field operator $a$ and the detuning $\Delta_{C,0} = \omega_P -\omega_C$. It is unoccupied in the absence of scattered pump photons in the longitudinal cavity mode with the mode function $g(\bm{r}) = g_C \cos{kx}$. 
The amplitude $g_C = \sqrt{U_0 \Delta_A}$ depends on the cavity frequency shift by a single atom $U_0$ and  $\Delta_A$. 
In addition, we include the s-wave scattering between the condensate atoms by the pseudopotential $U$ and consider the effective Hamiltonian with the atomic internal ground state field operators $\psi(\bm{r})$ in the form \cite{maschler2008} (with $\hbar=1$) 
\begin{align}
&H = -\Delta_{C,0} a^{\dagger} a + \int d\bm{r} \psi^{\dagger}(\bm{r}) H_A^{(1)} \psi(\bm{r}) \, , \nonumber \\
&H_A^{(1)} = - \frac{1}{2m} \nabla^2 + \frac{U}{2}\psi^{\dagger}(\bm{r}) \psi(\bm{r}) \nonumber \\
&+ \frac{1}{\Delta_A}\left[ h^2(\bm{r}) + g^2(\bm{r}) a^{\dagger} a + g(\bm{r})h(\bm{r})(a+a^{\dagger})\right].\label{oneAtomField}
\end{align}
Two types of light-matter coupling occur. First, the coupling between the occupation numbers induces a dynamic frequency shift 
 characterized by  $U_0$ and the periodicity $\pi/k$ along the cavity axis. 
Below the Dicke phase transition, this coupling is much smaller than the atomic recoil energy,  $U_0 \langle a^{\dagger} a \rangle \ll \omega_R = k^2/(2m)$ and we include its zeroth-order contribution $\Delta_C = -\Delta_{C,0} + U_0 N_C/2$.
The second type of coupling is due to the atoms scattering the pump photons into the cavity and vice versa. It thus depends on both the single atom frequency shift $U_0$ and the pump Rabi frequency $h_P$. 
The combination of the pump and cavity mode function yields a $2\pi/(\sqrt{2}k)$-periodicity along the diagonal of the cavity and the pump direction. 

{\em Mode expansion and Bogoliubov rotation. --} 
Focusing on the regime below the critical pump strength we expand the atomic field operators in the non-localized regime in terms of the two-mode Bloch functions with the momentum operators $b_{\bm{p}}$ and $c_{\bm{p}}$ \cite{konya2014damping}, i.e., 
\begin{align}
\psi(\bm{r}) = \frac{1}{\sqrt{V_{2D}}} \sum_{\bm{p} \in \mathcal{P}} e^{i\bm{p}\bm{r}}\left( b_{\bm{p}} + 2 \cos{kx} \cos{ky} \; c_{\bm{p}} \right) \, ,\nonumber \\
\mathcal{P} = \left\lbrace \begin{pmatrix}\frac{2\pi j_x}{L_x}, & \frac{2\pi j_y}{L_y} \end{pmatrix}^T \left| j_x,j_y \in \mathbb{Z} \wedge |p_x|,|p_y| < \frac{k}{2} \right. \right\rbrace \, , \label{Blochbands}
\end{align}
with  $V_{2D}=L_x L_y$ being the atomic condensate volume. 
This expansion is motivated by the special role of the zero-momentum condensate mode $b_0$ and the checkerboard mode $c \equiv c_0$ with momentum $\sqrt{2}k$ for the phase transition \cite{nagy2008self,baumann2010}. 
Commonly, s-wave scattering is weak, and a macroscopic number $n_0$ of atoms occupies the condensate mode $b_0$.
Therefore, we can rewrite the condensate operators in the spirit of Bogoliubov theory as the total atom number reduced by fluctuations, i.e., 
\begin{align}
b_0,b_0^{\dagger} = \sqrt{n_0} = \sqrt{N_C - c^{\dagger} c - \sum_{\bm{p} \neq 0} \left( b_{\bm{p}}^{\dagger} b_{\bm{p}} + c_{\bm{p}}^{\dagger} c_{\bm{p}} \right)} \, .
\end{align}
To generalize Bogoliubov theory, we include operators of $\bm{p} \neq 0$ up to quadratic order and keep all terms of at least ${\cal O}(\sqrt{N_C})$.
The latter condition corresponds to including those interaction processes which involve the condensate mode, i.e., among others Umklapp processes are neglected.
Then, we apply the Bogoliubov rotation
\begin{align}
b_{\bm{p}} \to  b_{\bm{p}}\cosh{\alpha_{b,\bm{p}}} + b_{-\bm{p}}^{\dagger} \sinh{\alpha_{b,\bm{p}}} \, , \nonumber \\
c_{\bm{p}} \to c_{\bm{p}} \cosh{\alpha_{c,\bm{p}}}  + c_{-\bm{p}}^{\dagger} \sinh{\alpha_{c,\bm{p}}}  \, .
\end{align}
The coefficients depend on the atomic interaction $nU$ with the atom density $n = N_C / V_{2D}$ as well as on the respective band $\omega_{b,\bm{p}}^\prime = \bm{p}^2/(2m)\equiv \omega_{\bm{p}}$ and $\omega_{c,\bm{p}}^\prime = \omega_{\bm{p}} +  k^2/m + h_P^2/(4\Delta_A) = \omega_{\bm{p}} + 2 \omega_R - \overline{\omega}_P$.
The last term with the pump strength $\overline{\omega}_P = -h_P^2/4\Delta_A$ follows from the $h^2(\bm{r})$ term in the two-dimensional field Hamiltonian Eq.\ (\ref{oneAtomField}). 
The coefficients read for $\nu=b,c$
$\tanh{2\alpha_{\nu,\bm{p}}} =  nU/(\omega_{\nu,\bm{p}}^\prime+nU)$. 
After the Bogoliubov rotation, the new quasiparticle frequencies are 
\begin{align}
\omega_{\nu,\bm{p}} = \sqrt{\left( \omega_{\nu,\bm{p}}^\prime \right)^2 + 2nU\omega_{\nu ,\bm{p}}^\prime} \, . \label{BogoliubovQuasiparticleFrequencies}
\end{align}
Finally, we abbreviate the coefficients as
\begin{align}
&\phi_{1\bm{p}} = \cosh{\alpha_{b,\bm{p}}}\cosh{\alpha_{c,\bm{p}}} \, ,	&\phi_{2\bm{p}} = \sinh{\alpha_{b,\bm{p}}}\sinh{\alpha_{c,\bm{p}}}  \, ,\nonumber \\
&\theta_{1\bm{p}} = \cosh{\alpha_{b,\bm{p}}}\sinh{\alpha_{c,\bm{p}}} \, ,	&\theta_{2\bm{p}} = \sinh{\alpha_{b,\bm{p}}}\cosh{\alpha_{c,\bm{p}}}  \, , \label{BogoliubovCoefficients}
\end{align}
and their combinations as $\phi_{\bm{p}} = \phi_{1\bm{p}} + \phi_{2\bm{p}}$ as well as $\theta_{\bm{p}} = \theta_{1\bm{p}} + \theta_{2\bm{p}}$.
Moreover, the zero-momentum coefficient  follows as $\phi_0 = \cosh{\alpha} - \sinh{\alpha}$ with $\alpha \equiv\alpha_{c,0}$.
By this, we obtain the effective Hamiltonian 
\begin{align}
H = \Delta_C a^{\dagger} a + \omega_0 c^{\dagger} c  + \lambda_0 (a+ a^{\dagger}) (c + c^{\dagger}) + H_{\beta} + H_{\text{int}} \nonumber\\
H_{\beta} = \sum_{\bm{p}\neq 0} \left( \omega_{b,\bm{p}} b_{\bm{p}}^{\dagger} b_{\bm{p}} + \omega_{c,\bm{p}} c_{\bm{p}}^{\dagger} c_{\bm{p}} \right)
\end{align}
with $\omega_0 = \omega_{c,0}$ and $\lambda_0 = \sqrt{N_C} \phi_0 \lambda$.
The interaction Hamiltonian
\begin{align}
&H_{\text{int}} = \lambda (a+a^{\dagger}) K + \eta \phi_0 (c+c^{\dagger}) K \label{interactionHamiltonian} \\&+ \eta (c \cosh{\alpha}  - c^{\dagger}\sinh{\alpha} ) \bar{K}^{\dagger} + \eta (c^{\dagger}\cosh{\alpha}  - c \sinh{\alpha} ) \bar{K} \nonumber
\end{align}
 describes the coupling of the cavity mode and the checkerboard mode to the atomic fluctuations with $\bm{p} \neq 0$.
Its coupling parameters are $\lambda = \sqrt{-U_0 \overline{\omega}_P}$ to the cavity and $\eta = 2\sqrt{N_C}U/V_{2D}$ to the checkerboard mode. The two combinations of fluctuation operators are summarized as $K$ and $\bar{K}$.
We distinguish between the terms describing Landau and Beliaev processes $K = K^L - K^B$ and $\bar{K} = -\bar{K}^L + \bar{K}^B$ with 
\begin{align}
K^{ L}&= \sum_{\bm{p}\neq0} \phi_{\bm{p}} \left[ b_{\bm{p}}^{\dagger} c_{\bm{p}} + {\rm h.c.} \right] , 
K^B = \sum_{\bm{p}\neq0} \theta_{\bm{p}} \left[ c_{-\bm{p}} b_{\bm{p}} + {\rm h.c.}\right]\, , \nonumber\\
\bar{K}^L &= \sum_{\bm{p}\neq0} \left[ \theta_{2\bm{p}} c_{\bm{p}}b_{\bm{p}}^{\dagger} + \theta_{1\bm{p}} c_{\bm{p}}^{\dagger} b_{\bm{p}} \right]\, , \nonumber\\
\bar{K}^B &= \sum_{\bm{p}\neq0} \left[ \phi_{1\bm{p}} c_{-\bm{p}} b_{\bm{p}} + \phi_{2\bm{p}}b_{\bm{p}}^{\dagger}c_{-\bm{p}}^{\dagger} \right] \, . 
\label{LandBprocesses}
\end{align}
When in Eq.\ (\ref{interactionHamiltonian}) an excitation  in the cavity or the checkerboard mode is created or annihilated, Landau processes describe the corresponding  annihilation (creation) of a phonon in one band and the corresponding creation (annihilation) in the other band with identical quasimomentum $\bm{p}$. Instead, Beliaev processes create or annihilate simultaneously a phonon in both bands with opposite $\bm{p}$. For both types, the Bogoliubov rotation 
 induces a momentum dependent coupling parameter $\phi_{\bm{p}}$ or $\theta_{\bm{p}}$.

{\em Landau and Beliaev quasiparticle damping.-- }
To reveal the influence of the quantum fluctuations on the system of the cavity and the checkerboard mode, we interpret the $\bm{p} \neq 0$ modes as a dissipative bath and employ the imaginary time-path integral formalism. It relies on the correlator $\langle H_{\text{int}}(\tau) H_{\text{int}}(\tau') \rangle_{\beta}$ \cite{napoli1994}.
The displacements of the system modes $q_C = (a+a^{\dagger})/\sqrt{2\Delta_C}$ and $q_A = (c+c^{\dagger})/\sqrt{2\omega_0}$ are used to describe its degrees of freedom, with 
 the influence functional $S_{\text{infl}}[q_C,q_A] = -\frac{1}{2}\int_0^{\beta} d\tau \int_0^{\beta} d\tau' \langle H_{\text{int}}(\tau) H_{\text{int}}(\tau') \rangle_{\beta}$	with the inverse temperature $\beta=1/T$ and with 
\begin{align}
\label{bathCorrelator}
\langle &H_{\text{int}}(\tau)  H_{\text{int}}(\tau') \rangle_{\beta} =  \kappa_C(\tau-\tau')\Delta_C q_C(\tau) q_C(\tau') \\
& +\kappa_{AC}(\tau-\tau') \sqrt{\Delta_C\omega_0}[q_C(\tau)q_A(\tau') + q_A(\tau) q_C(\tau') ] \nonumber\\
&+\kappa_{A}(\tau-\tau')\omega_0 q_A(\tau)q_A(\tau') 
+\kappa_{\dot{A}} (\tau-\tau') \frac{1}{\omega_0}\dot{q}_A(\tau)\dot{q}_A(\tau'). \nonumber
\end{align}
The correlators are sorted depending on the system modes they couple to, as shown in Fig.\ \ref{system Bath} 
(see Supplemental Material \cite{supplementary}). We label the ensuing dissipation channels by $x\in  \{ C, AC, A , \dot{A}\}.$
The kernels $\kappa_{x}(\tau)$  are related to the spectral densities $G_{x}(\omega)$ of the respective quasiparticle damping channel as
\begin{align}
\kappa_{x}(\tau) = \int_0 ^\infty d\omega \, G_{x}(\omega) D_\omega(\tau) \, ,
\end{align}
with the free thermal Green's function $D_\omega(\tau)  = [1+n(\omega)]e^{-\omega \tau}+n(\omega)e^{\omega \tau}$ with 
$n(\omega)$ being the Bose-Einstein distribution. The spectral density is composed of the two parts as  $ G_{x}(\omega)= G_{x}^L(\omega)\Theta(\omega_0-\omega)+G_{x}^B(\omega)\Theta(\omega-\omega_0)$, separated by the border $\omega_0$ of the frequency support (see below). Here, $\Theta(x)$ is the Heaviside function.
\\
\begin{figure}
\centering
\includegraphics[scale=.16]{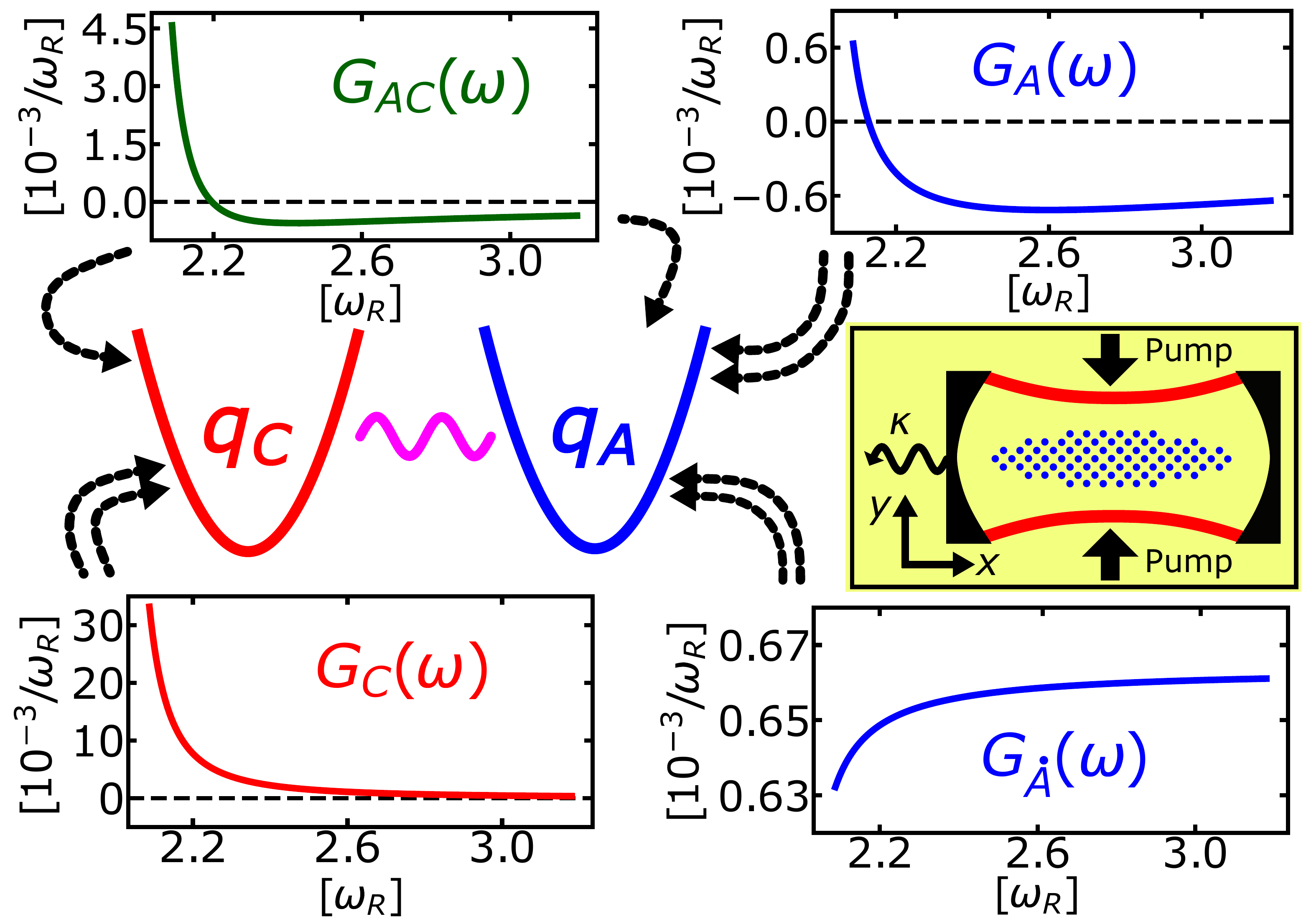}
\caption{Zero-temperature Beliaev spectral densities $G_{x}(\omega)$ describing the quasiparticle quantum dissipative baths acting on the cavity ($q_C$, red) and atom checkerboard ($q_A$, blue)  mode, respectively. 
The shaded plot shows a sketch of the cavity BEC setup with the identical color coding for the quantum occupation of the cavity mode and the checkerboard pattern atomic fluctuation in the cigar shaped BEC. In addition the loss through one of the cavity mirrors $\kappa$ is displayed which will be included in the theoretical model at a latter stage.
The parameters are $\overline{\omega}_P = 0.01 \omega_R$, $nU = 0.1 \omega_R$, $U_0 = -10^{-3} \omega_R$, $N_C = 5 \times 10^4$,  $L_x = 60$ \textmu m, $L_y = 11$ \textmu m,  $m = m_{\rm Rb87}$, and $\omega_R = 2\pi \times 3.56$ kHz. The spectral densities show strong sub-Ohmic forms and deviate from an Ohmic bath $G(\omega) \propto \omega$.}
\label{system Bath}
\end{figure}
The spectral densities follow in the discrete  form  as $G^{L/B}_{x}(\omega) = \sum_{\bm{p}\neq0} \gamma_{x} f_{x}^{L/B}(\omega_{\bm{p}}) \,\, \mathcal{N}_{\bm{p}}^{L/B} \delta(\omega - \omega_{\bm{p}}^{L/B})$, with the coupling parameters $(\gamma_{C},\gamma_{AC},\gamma_{A},\gamma_{\dot{A}}) = (\lambda^2,\lambda\eta\phi_0,\eta^2\phi_0^2/2,\eta^2/2\phi_0^2)$ and the combinations of Bose-Einstein distributions for Landau and Beliaev damping as 
$\mathcal{N}_{\bm{p}}^L = n(\omega_{b,\bm{p}})-n(\omega_{c,\bm{p}})$ and  $\mathcal{N}^B = 1 + n(\omega_{b,\bm{p}}) + n(\omega_{c,\bm{p}})$, respectively.
 The functions $f_{x}^{L/B}(\omega_{\bm{p}})$ are given in the Supplemental Material \cite{supplementary}. 

All Landau processes share the same finite frequency range $\omega_{\bm{p}}^L = \omega_{c,\bm{p}} - \omega_{b,\bm{p}} \in \left( 2 \omega_R - \overline{\omega}_P, \omega_0 \right)$.
The finite support results from the Brillouin zone of the phonon bands in Eq.\ (\ref{Blochbands}).
Beliaev processes exist in the range $\omega_{\bm{p}}^B = \omega_{c,\bm{p}} + \omega_{b,\bm{p}} \in \left( \omega_0 , 3\omega_R - \overline{\omega}_P + 2nU \right)$.
We find exact analytic expressions for the spectral densities of the dissipative Landau and Beliaev bath in the continuum limit in the form 
\begin{align}
G_{x}^{L/B}(\omega) = \frac{V_{2D}}{ 2\pi} m  \, \gamma_{x}  \,f_{x}^{L/B}(\omega)  \,\frac{ \mathcal{N}^{L/B}(\omega)}{|g_{L/B}'(\omega)|} \, .
\end{align}
The details of the derivation and the explicit forms of $f_{x}^{L/B}(\omega)$, $\mathcal{N}^{L/B}(\omega)$ and $g_{L/B}'(\omega)$ are given in the Supplemental Material \cite{supplementary}.
They encompass combinations of the coefficients in Eq.\ (\ref{BogoliubovCoefficients}) and the quasiparticle frequencies, Eq.\  (\ref{BogoliubovQuasiparticleFrequencies}). They depend on temperature,  the s-wave scattering $nU$ and the strength of the transversal pump $\overline{\omega}_P$ and can thus be tuned from outside. 

Landau processes are inherently thermal and vanish at zero temperature, such that the quantum fluctuations only include  Beliaev processes. The Beliaev spectral densities are shown in Fig.\ \ref{system Bath}. Two channels of the Beliaev damping, both the $\kappa_{AC}$ and, more pronounced, the $\kappa_{A}$ consist of several processes as follows from Eq.\ (\ref{bathCorrelator}). Evidently, processes compete with each other, thereby providing damping or anti-damping. This is reflected in the spectral densities with positive (damping) or negative (anti-damping) spectral weights in different frequency regions.  The possibility to tune the bath spectra 
 alters the cross-over frequency that separates damping from anti-damping.

The spectral densities diverge at the borderline of Landau and  Beliaev damping at 
$\omega = \omega_0$, where the excluded checkerboard mode ($|\bm{p}|=0$) lies. The thermal distributions $\mathcal{N}^{L/B}(\omega)$ have a pole of order $1$ at $\omega_0$. The term $f_{x}^{L/B}(\omega) / |g_{L/B}'(\omega)|$ does not diverge at this point.
Although being finite, 
$\omega=\omega_0$ is still a singular point forming a cusp. 
In accordance with their frequency ranges, the Landau functions $f^L/|g_L'|$ form the left side (for $\omega < \omega_0$)
and the Beliaev functions $f^B/|g_B'|$ the right side  (for $\omega > \omega_0$) of each cusp along the frequency axis.
We find that $\lim_{\omega \to \omega_0} \frac{d}{d\omega} f_x^{L/B}(\omega)/|g'_{L/B}(\omega)| \propto |\omega-\omega_0|^{-1/2}$ (for common parameters $\omega_{c,0}'\gg nU$) \cite{supplementary}. 
Importantly, it follows that the spectral function at zero temperature near the singular point $\omega_{0}$  is characterized by  $G_{x}^{B}(\omega)\propto (\omega-\omega_0)^{1/2}$ with the spectral exponent $s=1/2$.  This sub-Ohmic behavior has been introduced phenomenologically previously \cite{nagy2015nonequilibrium} and is now derived microscopically in the present work. Sub-Ohmic quantum fluctuations commonly induce strong, exotic non-Markovian real-time dynamics.  

{\em Quantum fluctuations in the atom and photon sectors.--}
With the imaginary-time influence functional, observables of the fluctuations can be calculated. The partition function $Z$
follows after a Matsubara expansion in $\beta$. We consider $\langle q_C^2 \rangle = - (\beta \Delta_C)^{-1} \partial_{\Delta_C} \ln{Z}$ for the cavity and $\langle q_A^2 \rangle = - (\beta \omega_0)^{-1} \partial_{\omega_0} \ln{Z}$ for the atomic sector. 
We additionally include a weak cavity photon loss by an Ohmic bath with damping constant $\kappa$, as determined experimentally \cite{nagy2011critical,brennecke2013}, and Drude-cut off $\omega_D$.
\begin{figure}
\centering
\includegraphics[scale=.5]{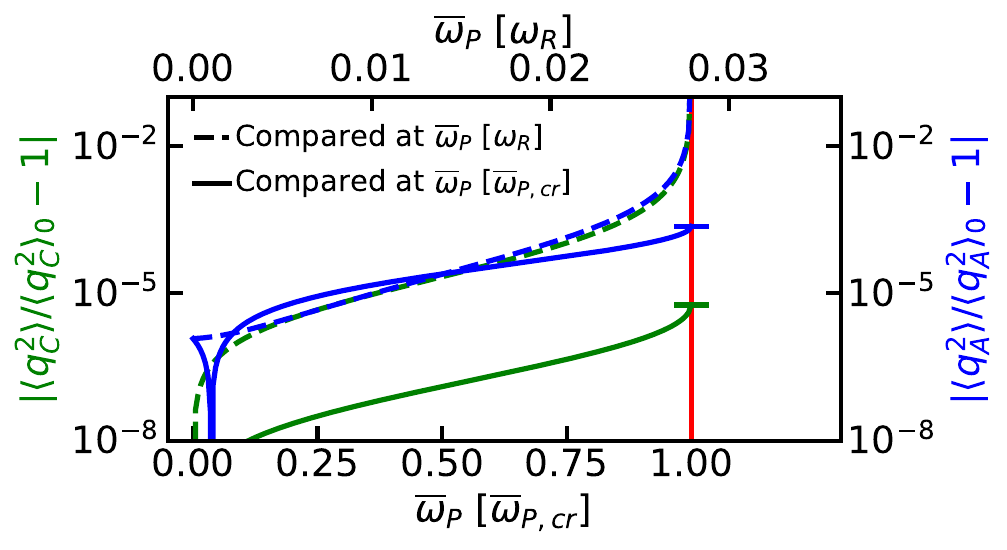}
\caption{Relative difference between the quantum fluctuations in the photon (green, left) and the atom (blue, right) sector with quasiparticle damping and without the quasiparticle damping for varying pump strengths. The upper axis is scaled to the recoil frequency and the result shown as dashed lines, thus, includes the effect of the Stokes shift. On the other hand, the lower axis is scaled in units of the critical pump strength and accordingly the result shown by the solid lines does not include the Stokes shift. The red vertical marks the critical point of the modified Dicke phase transition. The temperature  is $\beta \omega_R = 1.71  \times 10^3$ such that thermal fluctuations are negligible and quantum fluctuations prevail. Parameters for  
the recoil resolved experimental setup of Ref.\  \cite{kessler2014optomechanical,klinder2016bose} are used: 
 $\Delta_C = 2 \omega_R$, $nU = 1.6 \times 10^{-2} \omega_R$, $\kappa = 1.25 \omega_R$, $\omega_D = 10^9 \omega_R$, $n_{\text{max}} = 10^4$, remaining  parameters are as in Fig.\ \ref{system Bath}.  }
\label{omPTuneVariances}
\end{figure}
We investigate the strongly entangled system when approaching the critical point from below by increasing the atom-cavity coupling $\lambda = \sqrt{-U_0 \overline{\omega}_P}$.
In Fig.\  \ref{omPTuneVariances}, we compare the atom and cavity fluctuations with ($\langle \cdot \rangle$) and without ($\langle \cdot \rangle_0$) Landau and Beliaev damping. Close to the critical point $\overline{\omega}_{P,cr}$,  both relative differences diverge.  This behavior is generic also for thermal fluctuations. The fluctuations in the cavity and checkerboard modes differ near $\overline{\omega}_P=0$. Clearly, without pumping, the empty cavity does not couple to the atomic fluctuations. 
Naturally, the two dissipation channels $\kappa_{A}$ and $\kappa_{\dot{A}}$ are coupled to the checkerboard mode by the s-wave scattering $U$. 

{\em Role of the Stokes shifts.--} The presence of the quasiparticle dissipative baths induces a Stokes shift to the two-mode system, leading also to a shift of the critical point. This can be seen in Fig.\ \ref{omPTuneVariances}, where the fluctuations are finite exactly at the ``naked''  critical point. For these parameters, the relative deviation is  $2.3 \times 10^{-4}$ in the pump strength. The  Stokes shift can be derived analytically from the $n=0$ term in the Matsubara expansion of the observables.
We find
\begin{align}
\lambda_{cr}^2 = \frac{\Delta_C^2 + \kappa^2}{\Delta_C} \frac{\omega_0 - R_{A}}{\tilde{R}_C\left( \omega_0 - R_{A} \right)+( 2 \phi_0 \sqrt{N_C} - \tilde{R}_{AC} )^2}.
\label{criticalLambda}
\end{align}
The reorganization energies $R_{x}=\int_0^\infty d\omega G_{x}(\omega)$  shift the critical point.
The tilde indicates that the $\lambda$-dependence has been removed from the spectral densities involving the cavity sector before computing $R_{x}$. In an exact treatment, $R_{x}$ depend on the pumping, and, with them, their reorganization energy.  Similarly, $\Delta_C$ depends on the single atom-cavity shift $U_0$.
This renders Eq.\ (\ref{criticalLambda})  implicit. Without the quasiparticle bath,  $R_{x}=0$ the known critical point of an open Dicke phase transition 
 is recovered \cite{nagy2011critical}. 
In addition, 
the quantum fluctuations of the system with and without the atomic quasiparticle damping differ even if we correct for the Stokes shift. This can be seen as the difference between the solid and dashed lines in Fig.\ \ref{omPTuneVariances}.  However, the deviation saturates close to the critical point. The critical exponent is unchanged despite the sub-Ohmic characteristics of the quasiparticle baths. The sub-Ohmic signature is not relevant for the polariton soft mode exactly at the phase transition and, thus, does not imprint on the critical exponent \cite{nagy2015nonequilibrium,konya2018nonequilibrium}. 

\begin{figure}
\centering
\includegraphics[scale=.5]{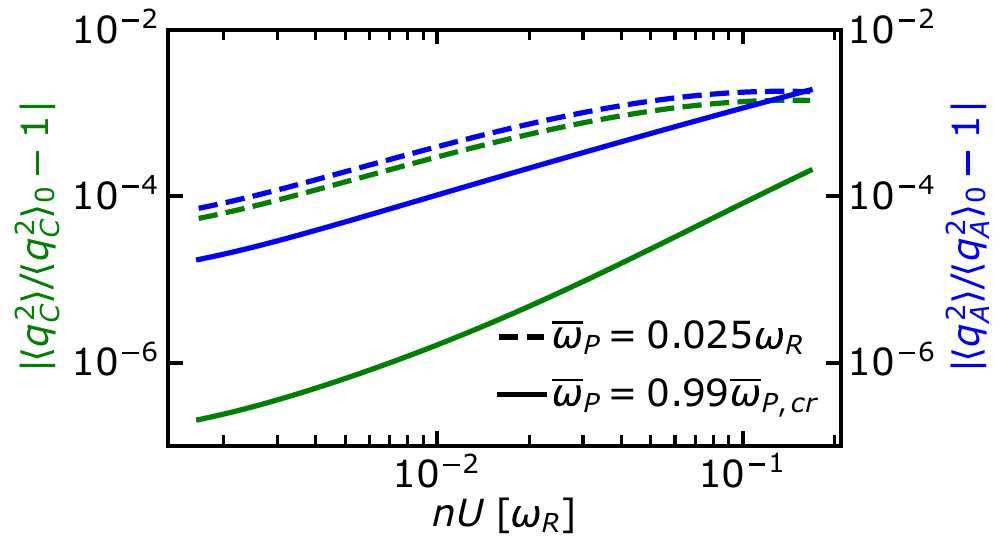}
\caption{Relative difference between the quantum fluctuations in the photon (green, left) and the atom (blue, right) sector in dependence of the atom-atom interaction $nU$ for two different values of  $\overline{\omega}_P$ and for the same parameters as in Fig.\ \ref{omPTuneVariances}.}
\label{nUtuneVariances}
\end{figure}

{\em Impact of s-wave scattering.--} Finally, the obvious channel to control the atomic quantum fluctuations is the  
s-wave scattering. The change of the quantum fluctuations for increasing  $U$ is displayed in Fig.\ \ref{nUtuneVariances}.
For low temperature, an increase in $nU$ already yields enhanced damping because the pump strength is held constant. Yet, the critical pump strength is shifted even without the bath, as seen from Eq.\  (\ref{criticalLambda}), due to the dependence of the checkerboard mode $\omega_0$ on $nU$, leading correspondingly to an increase of the necessary pump strength $\overline{\omega}_P$.  If we correct for the Stokes shift which grows approximately linear with $U$ \cite{supplementary}, the influence becomes more pronounced.
Additionally, with larger $nU$ the difference of influence seen on the cavity and checkerboard mode after correcting for the shift of the critical point is greatly reduced. In general, the strength of the fluctuations is tuned over two orders of magnitude.

{\em Conclusions.-- } Starting from a microscopic Hamiltonian, we have derived in closed analytic form the spectral characteristics of the quantum fluctuations of the atom and photon sector of a cavity BEC system in the presence of atomic s-wave scattering. 
The atomic quantum fluctuations show strong sub-Ohmic signatures of the quasiparticle damping, with the ensuing strongly non-Markovian dynamics. Furthermore, different channels of the Beliaev processes compete,  leading to a peculiar spectral behavior with competing damping and antidamping.  
We identify two channels to control and enhance the fluctuations, the atomic s-wave scattering and the strength of the transversal pump.  We also provide analytic expressions for the observables of the fluctuations and determine the Stokes shift of the critical atom-cavity coupling in the presence of quasiparticle damping. 
The controllability of the quantum fluctuations and access to the dynamics bring detailed understanding of sub-Ohmic quantum baths relevant not only in cavity BEC but furthermore in such fields as superconducting qubits, quantum information, nanomechanical and glassy systems, and quantum dots. In these systems the time-nonlocal dynamics can be exploited given an understanding of their spectral properties of which we provide with these results.

This work was supported by the Deutsche Forschungsgemeinschaft SFB 925 ``Light-induced dynamics and control of correlated quantum systems'' (170620586) within project C05, the UP System Balik PhD Program (OVPAA-BPhD-2021-04), the QuantERA II Programme that has received funding from the European Union’s Horizon 2020 research and innovation programme under Grant Agreement No.\ 101017733 and the Cluster of Excellence “Advanced Imaging of Matter” EXC 2056 (390715994). We acknowledge financial support from the Open Access Publication Fund of Universität Hamburg. We thank Axel Pelster and Milan Radonjic for helpful discussions.

\end{document}


\title{Supplemental Material to  \\[2mm] Enhancing exotic quantum fluctuations in strongly entangled cavity BEC systems}

\author{Leon Mixa}
\email{lmixa@physnet.uni-hamburg.de}
\affiliation{I. Institut für Theoretische Physik, Universität Hamburg, Notkestraße 9, 22607 Hamburg, Germany}
\author{Hans Ke{\ss}ler}
\affiliation{Zentrum für Optische Quantentechnologien and Institut für Laser-Physik, Universität Hamburg, 22761 Hamburg, Germany}
\author{Andreas Hemmerich}
\affiliation{Zentrum für Optische Quantentechnologien and Institut für Laser-Physik, Universität Hamburg, 22761 Hamburg, Germany}
\affiliation{The Hamburg Center for Ultrafast Imaging, Luruper Chaussee 149, 22761 Hamburg, Germany}
\author{Michael Thorwart}
\affiliation{I. Institut für Theoretische Physik, Universität Hamburg, Notkestraße 9, 22607 Hamburg, Germany}
\affiliation{The Hamburg Center for Ultrafast Imaging, Luruper Chaussee 149, 22761 Hamburg, Germany}

\begin{abstract}
In this Supplemental Material, we provide a summary of the derivation of the correlation functions and the ensuing spectral densities of the dissipative quasiparticle baths. In addition, we determine the critical exponents around the cusp singularity. Furthermore, we show the results of the atomic and photonic fluctuations at finite temperature and for a parameter set corresponding to a strongly damped cavity. Finally, we illustrate the shift of the critical point of the Dicke quantum phase transition due to atom-atom interaction.
\end{abstract}

\maketitle
\section{Bath Correlation Functions and Spectral Densities}
The Hamiltonian describing the interaction of the fluctuations with the system modes is (with $\hbar=1$) 
\begin{CAlign}
H_{\text{int}} = \lambda (a+a^{\dagger}) K + \eta \phi_0 (c+c^{\dagger}) K + \eta (c \cosh{\alpha} - c^{\dagger} \sinh{\alpha}) \bar{K}^{\dagger} + \eta (c^{\dagger} \cosh{\alpha} - c \sinh{\alpha}) \bar{K}\, .
\end{CAlign}
We proceed to find the influence functional of the quantum dissipative bath by imaginary time path integral methods.
To this end, the bath average
\begin{CAlign}
&\langle \cdot \rangle_{\beta} = \frac{1}{\mathcal{Z}_{\beta}} \int \mathcal{D}\left[ b^{\dagger},b,c^{\dagger},c\right] \left( \cdot \right) e^{-S_{\beta}\left[ b^{\dagger},b,c^{\dagger},c\right]}, \\
&H_{\beta} = \sum_{\bm{p} \neq 0} \left( \omega_{b,\bm{p}} b_{\bm{p}}^{\dagger} b_{\bm{p}} + \omega_{c,\bm{p}} c_{\bm{p}}^{\dagger} c_{\bm{p}} \right)
\end{CAlign}
is used.
Within this approach, the influence functional can be calculated in an exact form [24,\,33] and requires no more than the evaluation of the correlator
\begin{CAlign}
&\langle H_{\text{int}}(\tau) H_{\text{int}}(\tau') \rangle_{\beta} = \lambda^2 (a(\tau)+a^{\dagger}(\tau)) \langle K(\tau) K(\tau') \rangle_{\beta} (a(\tau')+a^{\dagger}(\tau')) \nonumber \\
&+ \lambda \eta \phi_0 (a(\tau)+a^{\dagger}(\tau)) \left[ \langle K(\tau) K(\tau') \rangle_{\beta} + \frac{1}{2} \langle K(\tau) \bar{K}^{\dagger}(\tau')\rangle_{\beta} + \frac{1}{2} \langle K(\tau) \bar{K}(\tau') \rangle_{\beta} \right]  (c(\tau') + c^{\dagger}(\tau')) \nonumber \\
&+ \eta \lambda \phi_0 (c(\tau) + c^{\dagger}(\tau)) \left[ \langle K(\tau) K(\tau') \rangle_{\beta} + \frac{1}{2} \langle \bar{K}^{\dagger}(\tau) K(\tau') \rangle_{\beta} + \frac{1}{2} \langle \bar{K}(\tau) K(\tau') \rangle_{\beta} \right] (a(\tau') + a^{\dagger}(\tau')) \nonumber \\
&+ \eta^2 \phi_0^2 (c(\tau)+c^{\dagger}(\tau)) \left[ \langle K(\tau) K(\tau') \rangle_{\beta} + \langle K(\tau) \bar{K}(\tau') \rangle_{\beta} + \langle \bar{K}(\tau) K(\tau') \rangle_{\beta} \right] (c(\tau') + c^{\dagger}(\tau') \nonumber \\
&+ \frac{\eta^2}{2} c(\tau) \left( \cosh^2{\alpha} + \sinh^2{\alpha} \right)  \left[ \langle \bar{K}^{\dagger}(\tau) \bar{K}(\tau') \rangle_{\beta} + \langle \bar{K}(\tau) \bar{K}^{\dagger}(\tau') \rangle_{\beta} \right]  c^{\dagger}(\tau') \nonumber \\
&+ \frac{\eta^2}{2} c^{\dagger}(\tau)  \left( \cosh^2{\alpha} + \sinh^2{\alpha} \right) \left[ \langle \bar{K}^{\dagger}(\tau) \bar{K}(\tau') \rangle_{\beta} + \langle \bar{K}(\tau) \bar{K}^{\dagger}(\tau') \rangle_{\beta} \right] c(\tau') \nonumber \\
&- 2 \eta^2 c(\tau) \cosh{\alpha} \sinh{\alpha} \langle \bar{K}(\tau) \bar{K}(\tau') \rangle_{\beta} c^{\dagger}(\tau') - 2 \eta^2  c^{\dagger}(\tau) \cosh{\alpha} \sinh{\alpha} \langle \bar{K}(\tau) \bar{K}(\tau') \rangle_{\beta} c(\tau') \nonumber \\
&- \eta^2 c(\tau)  \cosh{\alpha} \sinh{\alpha} \left[ \langle \bar{K}^{\dagger}(\tau) \bar{K}(\tau') \rangle_{\beta} + \langle \bar{K}(\tau) \bar{K}^{\dagger}(\tau') \rangle_{\beta} \right] c(\tau') \nonumber \\
&- \eta^2 c^{\dagger}(\tau) \cosh{\alpha} \sinh{\alpha} \left[ \langle \bar{K}^{\dagger}(\tau) \bar{K}(\tau') \rangle_{\beta} + \langle \bar{K}(\tau) \bar{K}^{\dagger}(\tau') \rangle_{\beta} \right] c^{\dagger}(\tau') \nonumber \\
& +\eta^2 c(\tau) \left( \cosh^2{\alpha} + \sinh^2{\alpha} \right) \langle \bar{K}(\tau) \bar{K}(\tau') \rangle_{\beta} c(\tau') \nonumber \\
&+ \eta^2 c^{\dagger}(\tau)  \left( \cosh^2{\alpha} + \sinh^2{\alpha} \right) \langle \bar{K}(\tau) \bar{K}(\tau') \rangle_{\beta} c^{\dagger}(\tau').
\end{CAlign}
We work in imaginary time and with stationary states. Therefore, the formalism inherently respects detailed balance, i.e., symmetry under $\tau \to  \beta - \tau$ by construction. 
The occurring correlations are evaluated in the interaction picture $b_{\bm{p}}(\tau) = e^{-\omega_{b,\bm{p}}\tau}b_{\bm{p}}$, $b_{\bm{p}}^{\dagger}(\tau)= b_{\bm{p}}^{\dagger} e^{+\omega_{b,\bm{p}}\tau}$ and $c_{\bm{p}}(\tau)$, $c_{\bm{p}}^{\dagger}(\tau)$, analogously.
In the bath averages, the mixing of the Landau and Beliaev processes yields zero contributions. 
Therefore, $\langle K(\tau) K(\tau') \rangle_{\beta} = \langle K^L(\tau) K^L(\tau') \rangle_{\beta}  + \langle K^B(\tau) K^B(\tau')\rangle_{\beta} $ and analogous for all combinations of $K,\bar{K}$,  such that
\begin{CAlign}
\kappa_1^L(\tau-\tau') = & \langle K^L(\tau) K^L(\tau') \rangle_{\beta}  = \sum_{\bm{p} \neq 0}  \phi_{\bm{p}}^2 \left[ n(\omega_{b,\bm{p}}) - n(\omega_{c,\bm{p}}) \right] D_{\omega^L_{\bm{p}}}(\tau - \tau'), \nonumber \\
\kappa_1^B(\tau-\tau') =  &\langle K^B(\tau) K^B\tau') \rangle_{\beta}  = \sum_{\bm{p} \neq 0} \theta_{\bm{p}}^2 \left[ 1 + n(\omega_{b,\bm{p}}) + n(\omega_{c,\bm{p}}) \right] D_{\omega^B_{\bm{p}}}(\tau-\tau'),  
\end{CAlign}
with $n(\omega)$ being the Bose-Einstein distribution and with $D_{\omega}(\tau)$ being the free thermal Green's function as defined in the main text. Furthermore,  we have the kernels 
\begin{CAlign}
\overrightarrow{\kappa}_2(\tau - \tau') = &-\langle K(\tau) \bar{K}(\tau') \rangle_{\beta}  = - \langle \bar{K}^{\dagger}(\tau) K(\tau') \rangle_{\beta}  = \overleftarrow{\kappa}_2( \beta - (\tau - \tau')), \nonumber \\
\overleftarrow{\kappa}_2(\tau - \tau') = &-\langle \bar{K}(\tau) K(\tau') \rangle_{\beta}  = -\langle K(\tau) \bar{K}^{\dagger}(\tau') \rangle_{\beta} , \nonumber \\
\kappa_2^L(\tau - \tau') := &\frac{1}{2}  \left[ \overrightarrow{\kappa}_2^L(\tau - \tau') + \overleftarrow{\kappa}_2^L(\tau-\tau') \right] 
= \sum_{\bm{p} \neq 0}  \frac{\phi_{\bm{p}} \theta_{\bm{p}}}{2}  \left[ n(\omega_{b,\bm{p}}) - n(\omega_{c,\bm{p}}) \right] D_{\omega^L_{\bm{p}}}(\tau-\tau'), \nonumber \\
\kappa_2^B(\tau - \tau') := &\frac{1}{2}  \left[ \overrightarrow{\kappa}_2^B(\tau - \tau') + \overleftarrow{\kappa}_2^B(\tau-\tau') \right]=
\sum_{\bm{p} \neq 0}  \frac{\phi_{\bm{p}} \theta_{\bm{p}}}{2}  \left[ 1 + n(\omega_{b,\bm{p}}) + n(\omega_{c,\bm{p}}) \right] D_{\omega^B_{\bm{p}}}(\tau-\tau'), 
\end{CAlign}
where we have enforced detailed balance. In addition, we have analogously 
\begin{CAlign}
\overrightarrow{\kappa}_3(\tau - \tau') = &\langle \bar{K}^{\dagger}(\tau) \bar{K}(\tau') \rangle_{\beta}  = \overleftarrow{\kappa}_3( \beta - (\tau - \tau')), \nonumber \\
\overleftarrow{\kappa}_3(\tau - \tau') = &\langle \bar{K}(\tau) \bar{K}^{\dagger}(\tau') \rangle_{\beta}, \nonumber \\
\kappa_3^L(\tau - \tau') := &\frac{1}{2} \left[ \overrightarrow{\kappa}_3^L(\tau-\tau') + \overleftarrow{\kappa}_3^L(\tau-\tau') \right] =
\sum_{\bm{p} \neq 0} \frac{\theta_{1\bm{p}}^2 + \theta_{2\bm{p}}^2}{2} \left[ n(\omega_{b,\bm{p}}) - n(\omega_{c,\bm{p}}) \right] D_{\omega^L_{\bm{p}}}(\tau -\tau'),  \nonumber \\
\kappa_3^B(\tau - \tau') := &\frac{1}{2} \left[ \overrightarrow{\kappa}_3^B(\tau-\tau') + \overleftarrow{\kappa}_3^B(\tau-\tau') \right]=
\sum_{\bm{p} \neq 0} \frac{\phi_{1\bm{p}}^2 + \phi_{2\bm{p}}^2}{2} \left[ 1 + n(\omega_{b,\bm{p}}) + n(\omega_{c,\bm{p}}) \right] D_{\omega^B_{\bm{p}}}(\tau - \tau'), 
\end{CAlign}
and 
\begin{CAlign}
\kappa_4^L(\tau - \tau') = &\langle \bar{K}^L(\tau) \bar{K}^L(\tau') \rangle_{\beta}  = \langle \bar{K}^{L\dagger}(\tau) \bar{K}^{L\dagger}(\tau') \rangle_{\beta} =
\sum_{\bm{p} \neq 0}  \theta_{1\bm{p}} \theta_{2\bm{p}} \left[ n(\omega_{b,\bm{p}}) - n(\omega_{c,\bm{p}}) \right] D_{\omega^L_{\bm{p}}}(\tau-\tau'), \nonumber \\
\kappa_4^B(\tau - \tau') &= \langle \bar{K}^B(\tau) \bar{K}^B(\tau') \rangle_{\beta}  = \langle \bar{K}^{B\dagger}(\tau) \bar{K}^{B\dagger}(\tau') \rangle_{\beta} =
\sum_{\bm{p} \neq 0}  \phi_{1\bm{p}} \phi_{2\bm{p}} \left[ 1 +n(\omega_{b,\bm{p}}) + n(\omega_{c,\bm{p}}) \right] D_{\omega^B_{\bm{p}}}(\tau -\tau'). 
\end{CAlign}
We recombine the different correlators now depending on the way how they couple to the system and obtain 
\begin{CAlign}
\langle H_{\text{int}}(\tau) H_{\text{int}}(\tau') \rangle_{\beta}  &= \Delta_C q_C(\tau) \kappa_C(\tau-\tau') q_C(\tau')  + \sqrt{\Delta_C \omega_0} q_C(\tau) \kappa_{AC}(\tau - \tau') q_A(\tau') \nonumber \\
&+ \sqrt{\Delta_C \omega_0}  q_A(\tau) \kappa_{AC}(\tau - \tau') q_C(\tau') + \omega_0 q_A(\tau) \kappa_{A}(\tau - \tau') q_A(\tau') \nonumber \\
&+ \frac{1}{\omega_0} \dot{q}_A(\tau) \kappa_{\dot{A}}(\tau-\tau') \dot{q}_A(\tau'), 
\end{CAlign}
with
\begin{CAlign}
\kappa_C(\tau-\tau') & = 2 \lambda^2 \kappa_1(\tau - \tau'), \nonumber \\
\kappa_{AC}(\tau-\tau') &= 2 \lambda \eta \phi_0 \left[  \kappa_1(\tau - \tau')-\kappa_2(\tau-\tau') \right], \nonumber \\
\kappa_{A}(\tau-\tau') &= \eta^2 \phi_0^2 \left[ 2\kappa_1(\tau - \tau') + \kappa_3(\tau - \tau') + \kappa_4(\tau-\tau') - 4 \kappa_2(\tau-\tau') \right], \nonumber \\
\kappa_{\dot{A}}(\tau-\tau') &= \frac{\eta^2}{\phi_0^2} \left[ \kappa_3(\tau-\tau') - \kappa_4(\tau-\tau') \right].  \label{kernels}
\end{CAlign}
\newpage
To obtain, the spectral densities, the kernels (\ref{kernels}) are now expressed as a frequency integral of their characterizing spectral density $G_x(\omega)$ and the free thermal Green's function, as shown in Eq.\ (11) in the main text. The spectral density is composed of the two parts as  $G_{x}(\omega)= G_{x}^L(\omega)\Theta(\omega_0-\omega)+G_{x}^B(\omega)\Theta(\omega-\omega_0)$, separated by the border $\omega_0$ of the frequency support (see discussion in the main text). Here, $\Theta(x)$ is the Heaviside function. 
We note that in the discrete $\bm{p}$-sum, the Heaviside function is not necessary, but needs to be added in the continuum version of the spectral densities. 
We find
\begin{CAlign}
G_C^L(\omega) = \sum_{\bm{p} \neq 0} 2 \lambda^2  \phi_{\bm{p}}^2 \mathcal{N}^L_{\bm{p}} \delta(\omega - \omega^L_{\bm{p}}), \nonumber \\
G_C^B(\omega) = \sum_{\bm{p} \neq 0} 2 \lambda^2  \theta_{\bm{p}}^2 \mathcal{N}^B_{\bm{p}} \delta(\omega - \omega^B_{\bm{p}}), \nonumber \\
G_{AC}^L(\omega) = \sum_{\bm{p} \neq 0} \lambda \eta \phi_0 \left( 2 \phi^2_{\bm{p}}-\phi_{\bm{p}} \theta_{\bm{p}} \right) \mathcal{N}^L_{\bm{p}} \delta(\omega - \omega^L_{\bm{p}}), \nonumber \\
G_{AC}^B(\omega) = \sum_{\bm{p} \neq 0} \lambda \eta \phi_0 \left( 2 \theta^2_{\bm{p}}-\phi_{\bm{p}} \theta_{\bm{p}} \right)  \mathcal{N}^B_{\bm{p}} \delta(\omega - \omega^B_{\bm{p}}), \nonumber \\
G_{A}^L(\omega) = \sum_{\bm{p} \neq 0} \eta^2 \phi_0^2 \frac{5 \phi^2_{\bm{p}}-4 \phi_{\bm{p}} \theta_{\bm{p}}-1}{2}  \mathcal{N}^L_{\bm{p}} \delta(\omega - \omega^L_{\bm{p}}), \nonumber \\
G_{A}^B(\omega) = \sum_{\bm{p} \neq 0} \eta^2 \phi_0^2 \frac{5 \theta^2_{\bm{p}}-4 \phi_{\bm{p}} \theta_{\bm{p}}+1}{2}  \mathcal{N}^B_{\bm{p}} \delta(\omega - \omega^B_{\bm{p}}), \nonumber \\
G_{\dot{A}}^L(\omega) = \sum_{\bm{p} \neq 0} \frac{\eta^2}{\phi_0^2}  \frac{(\theta_{1\bm{p}}-\theta_{2\bm{p}})^2}{2} \mathcal{N}^L_{\bm{p}} \delta(\omega - \omega^L_{\bm{p}}),  \nonumber \\
G_{\dot{A}}^B(\omega) = \sum_{\bm{p} \neq 0}\frac{\eta^2}{\phi_0^2}  \frac{(\phi_{1\bm{p}}-\phi_{2\bm{p}})^2}{2} \mathcal{N}^B_{\bm{p}} \delta(\omega - \omega^B_{\bm{p}})  \, , 
\end{CAlign}
with the combinations of Bose-Einstein distributions
\begin{CAlign}
\mathcal{N}^L_{\bm{p}} = n(\omega_{b,\bm{p}}) - n(\omega_{c,\bm{p}}), &	&\mathcal{N}^B_{\bm{p}} = 1 +  n(\omega_{b,\bm{p}}) + n(\omega_{c,\bm{p}}).
\end{CAlign}
From these equations, we read off the general structure
\begin{CAlign}
G^{L/B}_{x}(\omega) = \sum_{\bm{p}\neq0} \gamma_{x} f_{x}^{L/B}(\omega_{\bm{p}}) \,\, \mathcal{N}_{\bm{p}}^{L/B} \delta(\omega - \omega_{\bm{p}}^{L/B}) \, , 
\end{CAlign}
with the coupling parameters 
\begin{CAlign}
(\gamma_{C},\gamma_{AC},\gamma_{A},\gamma_{\dot{A}}) = (\lambda^2,\lambda\eta\phi_0,\eta^2\phi_0^2/2,\eta^2/2\phi_0^2) \, ,
\end{CAlign}
and
\begin{CAlign}
f^B_{C}(\omega_{\bm{p}}) &= 2 \theta_{\bm{p}}^2	 \, ,& f^L_{C}(\omega_{\bm{p}}) &= f^B_{C}(\omega_{\bm{p}}) + 2 \, , \nonumber \\
f^B_{AC}(\omega_{\bm{p}}) &= \theta_{\bm{p}} (2\theta_{\bm{p}} - \phi_{\bm{p}})	 \, ,&	f^L_{AC}(\omega_{\bm{p}}) &= f^B_{AC}(\omega_{\bm{p}}) + 2  \, , \nonumber\\
f^B_{A}(\omega_{\bm{p}}) &= 5\theta_{\bm{p}}^2 - 4 \phi_{\bm{p}} \theta_{\bm{p}} + 1  \, ,&	f^L_{A}(\omega_{\bm{p}}) &= f^B_{A}(\omega_{\bm{p}}) + 3  \, , \nonumber \\
f^B_{\dot{A}}(\omega_{\bm{p}}) &= (\phi_{1\bm{p}} - \phi_{2\bm{p}})^2	 \, ,&	f^L_{\dot{A}}(\omega_{\bm{p}}) &= (\theta_{1\bm{p}} - \theta_{2\bm{p}})^2 \, .
\label{characteristics}
\end{CAlign}

The continuum limit is introduced by  $\sum_{\bm{p} \neq 0} \to \frac{V_{2D}}{(2\pi)^2} \int_0^{k/\sqrt{2}} d\rho \, \rho \int_0^{2\pi} d\varphi$ based on the lattice formed by the periodic potential given in Eq.\ (4) in the main paper.
The arguments of the Dirac delta distributions controlling the frequency spectrum are 
\begin{CAlign}
g_L(\omega_{\rho}) = \omega - \omega^L(\omega_{\rho}), &	&g_B(\omega_{\rho}) = \omega - \omega^B(\omega_{\rho})\, ,
\end{CAlign}
where $\omega_{\rho} = \frac{\rho^2}{2m}$.
Then, the roots defined by $g_{L/B}(\omega_{\rho}=W(\omega))=0 $ lying in the respective frequency range for $g_{L/B}$ are 
\begin{CAlign}
W(\omega) = -\frac{\omega_{c,0}' + 2nU}{2} + \frac{\omega}{2} \sqrt{1 + \frac{(2nU)^2}{\omega^2 - \omega_{c,0}^{'2}}}\, .
\end{CAlign}
\newpage
We can next perform the $k$-space integral and find exact analytic expressions of our spectral densities in the form 
\begin{CAlign}\label{specdens}
G_{x}^{L/B}(\omega) = \frac{V_{2D}}{ 2\pi} m  \, \gamma_{x}  \,f_{x}^{L/B}(\omega)  \,\frac{ \mathcal{N}^{L/B}(\omega)}{|g_{L/B}'(\omega)|} \, , 
\end{CAlign}
with 
\begin{CAlign}
f_C^L(\omega) &= 2 \phi^2(W(\omega)) \, , &  
f_C^B(\omega) &= 2 \theta^2(W(\omega)) = f_C^L(\omega)-2 \, , \nonumber \\
f_{AC}^L(\omega) &= 2\phi^2(W(\omega))-\phi(W(\omega))\theta(W(\omega)) \, , &  
f_{AC}^B(\omega) &= 2\theta^2(W(\omega))-\phi(W(\omega))\theta(W(\omega)) \, , \nonumber \\
f_{A}^L(\omega) &= 5\phi^2(W(\omega))-4\phi(W(\omega))\theta(W(\omega))-1 \, , & 
f_{A}^B(\omega) &= 5\theta^2(W(\omega))-4\phi(W(\omega))\theta(W(\omega))+1 \, , \nonumber \\ 
f_{\dot{A}}^L(\omega) &= \left[ \theta_1(W(\omega)) - \theta_2(W(\omega)) \right]^2 \, ,  & 
f_{\dot{A}}^L(\omega) &= \left[ \phi_1(W(\omega)) - \phi_2(W(\omega)) \right]^2 \, ,
\end{CAlign}
with 
\begin{CAlign}
\theta(\omega) = \theta_1(\omega)+\theta_2(\omega) \, , && \phi(\omega) = \phi_1(\omega)+\phi_2(\omega) \, .
\end{CAlign}
Here, we have defined
\begin{CAlign}
\theta_{1}(\omega) & = \cosh \alpha_{b}(\omega)  \sinh \alpha_{c}(\omega) \, , & 
\theta_{2}(\omega) & =   \sinh \alpha_{b}(\omega) \cosh \alpha_{c}(\omega) \, ,\nonumber \\
\phi_1(\omega) &= \cosh \alpha_{b}(\omega) \cosh \alpha_{c}(\omega) \, , &
 \phi_2(\omega) & = \sinh \alpha_{b}(\omega) \sinh \alpha_{c}(\omega) \, , 
\end{CAlign}
with
\begin{CAlign}
\alpha_{b}(\omega) =\frac{1}{2} \artanh \left( \frac{nU}{W(\omega)+nU} \right)\, , &  &
\alpha_{c}(\omega) =\frac{1}{2} \artanh \left( \frac{nU}{W(\omega)+\omega_{c,0}'+nU}\right)\, . 
\end{CAlign}
Furthermore, we need
\begin{CAlign}
\mathcal{N}^L(\omega) = n(\omega_b(W(\omega))) - n(\omega_c(W(\omega))) \, , & & 
\mathcal{N}^B(\omega) = 1 + n(\omega_b(W(\omega))) + n(\omega_c(W(\omega))) \, , 
\end{CAlign}
with 
\begin{CAlign}
\omega_b(\omega) = \sqrt{W(\omega) [W(\omega)+2nU]}\, , & & 
\omega_c(\omega) = \sqrt{[W(\omega)+\omega_{c,0}'] [W(\omega)+\omega_{c,0}'+2nU]}\, .
\end{CAlign}
Finally, in Eq.\ (\ref{specdens}), we have defined 
\begin{CAlign}
g_{L/B}'(\omega) = - \left. \frac{d}{dy} \omega^{L/B}(y) \right|_{y=W(\omega)} \, .
\end{CAlign}

\section{Critical Exponents around the Cusp}
By analyzing the characteristic functions $f_{x}(W) /  |g'(W)|$, we can characterize the singularity at $\omega_0$.
Near the critical point defined by $W=0$, we find 
\begin{CAlign}
&\frac{d}{dW} \frac{f_C^B(W)}{|g_B'(W)|} \sim -\frac{a_C}{\sqrt{W}} + b_C, & &a_C = \frac{1}{\sqrt{2nU}} \left( 1 + \frac{(\omega_{c,0}' + nU) (\omega_{c,0}'+2nU)}{\omega_0^2}\right), \nonumber \\
& & &b_C = \frac{1}{\omega_0} \left( \frac{\omega_{c,0}'}{nU} + 1 \right) - \frac{(\omega_{c,0}'+nU)(\omega_{c,0}'+2nU)}{\omega_0^3}, \nonumber \\
&\frac{d}{dW} \frac{f_{AC}^B(W)}{|g_B'(W)|} \sim -\frac{a_{AC}}{\sqrt{W}} + b_{AC}, & &a_{AC} = \frac{1}{\sqrt{2nU}} \left( 1 + \frac{(\omega_{c,0}'+2nU)(\omega_{c,0}'+nU)}{2\omega_0^2} \right), \nonumber \\
& & &b_{AC} = \frac{\omega_{c,0}'}{2\omega_0nU} - \frac{(\omega_{c,0}'+2nU)(\omega_{c,0}'+nU)}{2\omega_0^3}, \nonumber \\
&\frac{d}{dW} \frac{f_{A}^B(W)}{2|g_B'(W)|} \sim -\frac{a_{A}}{\sqrt{W}} + b_{A}, & &a_{A} = \frac{1}{4\sqrt{2nU}} \left( 3 + \frac{(\omega_{c,0}'+nU)(\omega_{c,0}'+2nU)}{\omega_0^2} \right), \nonumber \\
& & &b_{A} = \frac{1}{2\omega_0}\left( \frac{\omega_{c,0}'}{nU} -1 \right) - \frac{(\omega_{c,0}'+2nU)(\omega_{c,0}'+nU)}{4\omega_0^3}, \nonumber \\
&\frac{d}{dW} \frac{f_{\dot{A}}^B(W)}{2|g_B'(W)|} \sim \frac{a_{\dot{A}}}{\sqrt{W}} - b_{\dot{A}}, & &a_{\dot{A}} = \frac{1}{4\sqrt{2nU}} \left( 1 - \frac{\omega_{c,0}' (\omega_{c,0}'+nU)}{\omega_0^2} \right), \nonumber \\
& & &b_{\dot{A}} = \frac{1}{4\omega_0} \left( \frac{\omega_{c,0}'}{nU}-1 \right) + \frac{\omega_{c,0}' (\omega_{c,0}' + nU)}{4\omega_0^3}.
\end{CAlign}
We can, therefore, approximate the spectral densities near $W = 0$ by the antiderivative with the constants determined by the respective characteristic value at $W=0$: $f_{x}(0)/|g'(0)|$. By this, we obtain 
\begin{CAlign}
&\frac{f_C^B(W)}{|g_B'(W)|} \approx - 2 a_C \sqrt{W} + b_C W + C_C	, &	&C_C = \frac{\omega_{c,0}'+2nU}{\omega_0}, \nonumber \\
&\frac{f_{AC}^B(W)}{|g_B'(W)|} \approx - 2 a_{AC} \sqrt{W} + b_{AC} W + C_{AC}	, &	&C_{AC} = \frac{\omega_{c,0}'+2nU}{2\omega_0}, \nonumber \\
&\frac{f_{A}^B(W)}{2|g_B'(W)|} \approx - 2 a_{A} \sqrt{W} - b_{A} W + C_{A},	&	&C_{A} = \frac{\omega_{c,0}'+2nU}{4\omega_0}, \nonumber \\
&\frac{f_{\dot{A}}^B(W)}{2|g_B'(W)|} \approx + 2 a_{\dot{A}} \sqrt{W} - b_{\dot{A}} W + C_{\dot{A}}, 	&	&C_{\dot{A}} = \frac{\omega_{c,0}'}{4\omega_0}.
\end{CAlign}
An analogous analysis for the characteristic functions of the Landau damping yields the left flanks of the cusps. However, due to the vanishing of Landau damping at $T \to 0$, this is physically irrelevant.

\begin{figure}
\centering
\includegraphics[scale=.7]{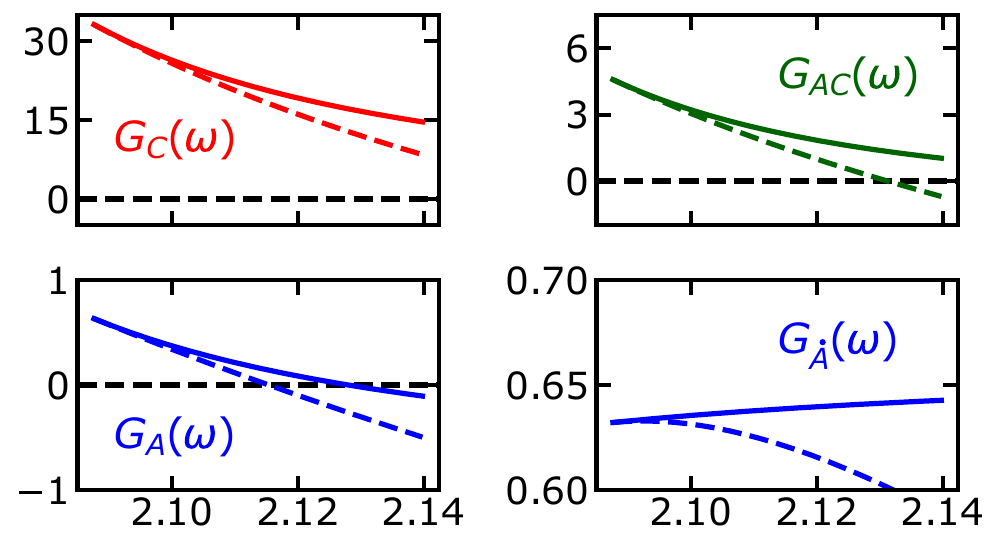}
\caption{\label{FS1}Comparison of the approximated spectral densities near the critical point at  zero temperature (dashed lines) and the exact ones (solid lines). The latter are identical to those shown in Fig.\ 1 in the main text. }
\end{figure}

We show the approximated spectral densities near the critical point at  zero temperature in comparison with the full result in Fig.\ \ref{FS1} for the parameters of Fig.\ 1 of the main text.
\newpage
\section{Observables in other parameter regimes}

In the main text, we have addressed mainly the case of pure quantum fluctuations at zero temperature. Furthermore, we have chosen a parameter set describing the case of a weakly damped optical cavity as realized in the Hamburg experiment. The approach introduced in our work is also valid at finite temperature and for strongly damped cavities, as, e.g., realized in the ETH Zürich experiment. 

\begin{figure}[h]
\centering
\includegraphics[scale=.5]{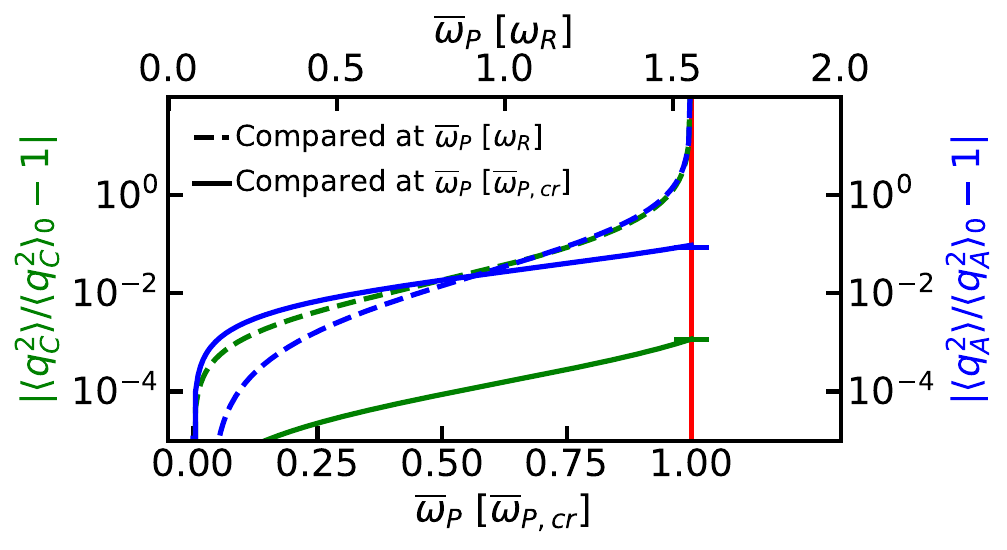} \hfill 
\includegraphics[scale=.5]{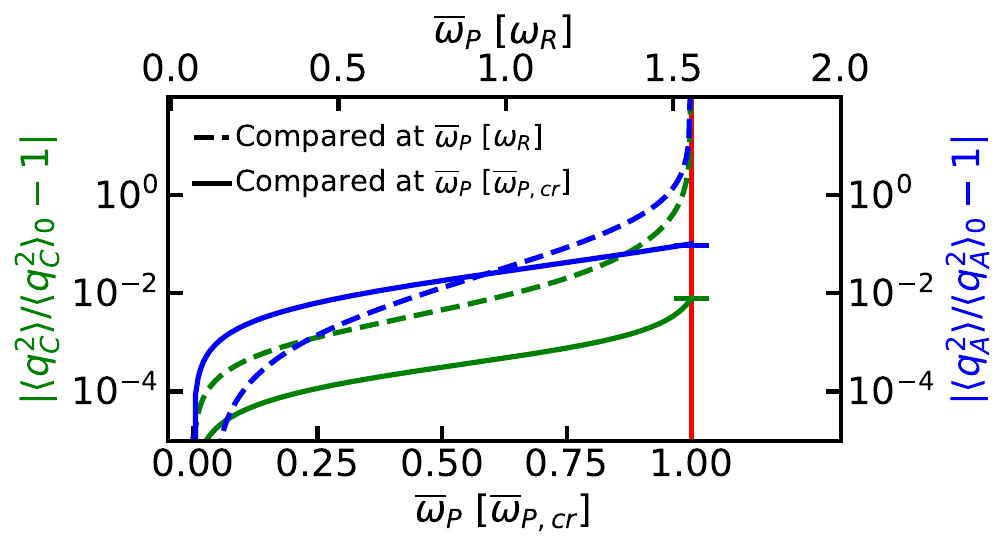}
\caption{Same results as in Fig.\ 2 of the main text, but for a finite temperature, as typical in an experiment where thermal fluctuations are dominant, i.e.,  $\beta \omega_R = 1.71$.  Left: All other parameters are the same as in Fig.\ 2 of the main text.  The relative shift of the critical point for this parameter set is $8.7 \times 10^{-2}$.  Right: Parameters are the same except for the cavity detuning $\Delta_C = 20 \omega_R$. The relative shift of the critical point is here $8.0 \times 10^{-2}$. \label{figS2}}
\end{figure}

Below, we show results for the quantum statistical fluctuations of the atom and the photon sector for a case of finite temperature and for stronger damped cavity. In Fig.\ \ref{figS2}, we show the same results as in Fig.\ 2 of the main text, but for a finite temperature of 
$\beta \omega_R = 1.71$. This is a typical configuration in an experiment where thermal fluctuations are dominant.  The left panel depicts the results when all other parameters are the same as  Fig.\ 2 of the main text, while the right panel shows the results for a different  cavity detuning $\Delta_C = 20 \omega_R$. 

\begin{figure}[h]
\centering
\includegraphics[scale=.5]{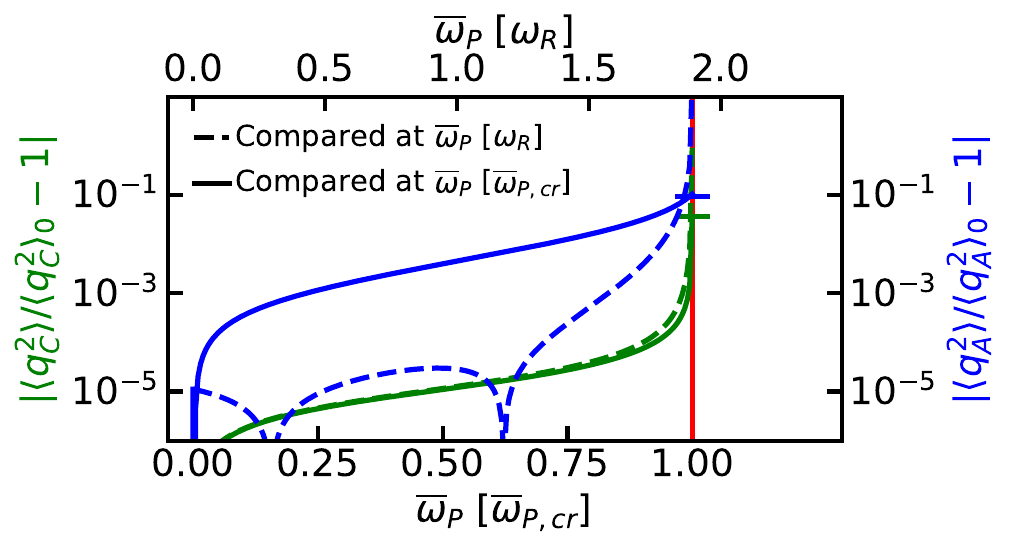}
\caption{Same results as in Fig.\ 2 of the main text, but for cavity parameters in the non-recoil resolved (overdamped) regime and for a  typical experimental temperature, i.e., $\Delta_C = 2000 \omega_R$, $\kappa = 1250 \omega_R$ and $\beta \omega_R = 1.71$. The resulting  Stokes shift is $2.6 \times 10^{-3}$, which is significantly smaller than  for the recoil-resolved regime. \label{figS3}}
\end{figure}

In our approach, we do take into explicit account the dynamics of the cavity light field (in imaginary time) instead of integrating over it. We thus may also consider the case of a strongly damped cavity and its fluctuations. In Fig.\  \ref{figS3}, we show the quantum statistical fluctuations for the case of a strongly damped cavity at finite temperature. In this case, we find a significant difference of the atomic fluctuations at absolute pump values. 


\section{Stokes shift of the critical point due to interaction}
\begin{figure}[h]
\centering
\includegraphics[scale=.5]{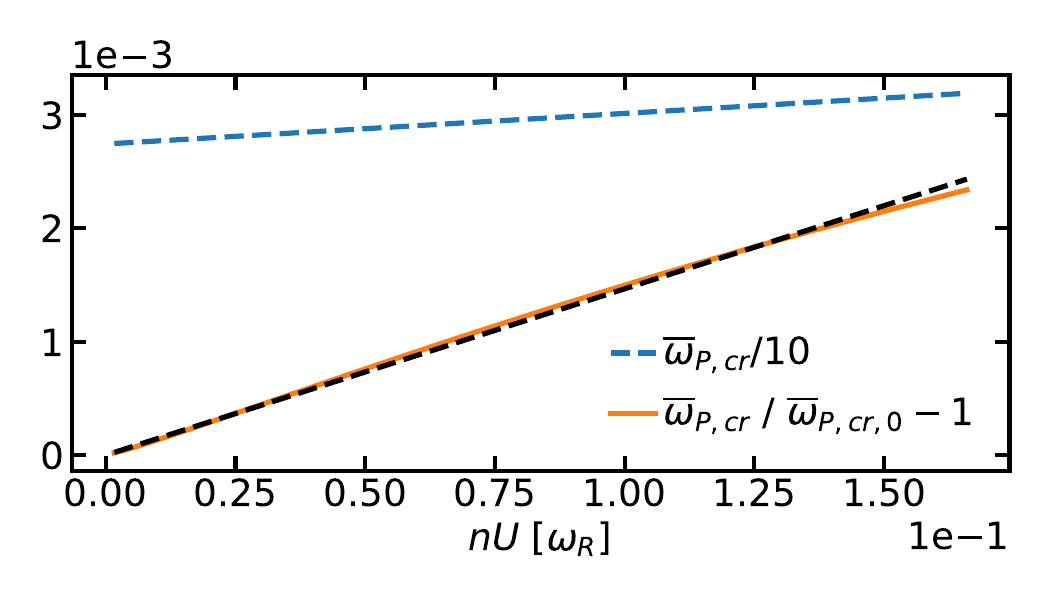}
\caption{Shift of the critical point (blue,dashed) and its deviation due to the Stokes shift (solid, orange) of the quantum phase transition as determined by the atom-atom interaction $U$. The black dashed line shows a linear fit to the Stokes shift growth. Parameters are the same as in Fig.\ 2 and 3 of the main text. \label{figS4}}
\end{figure}

The quasiparticle baths induce a Stokes shift to the polariton system which shifts also the critical point of the phase transition. The dependence of critical point on the atom-atom interaction $U$, which determines the Stokes shift as follows from Eq.\ (13) of the main paper, is shown in Fig.\ \ref{figS4}.
\nocite{*}